\documentclass[11pt]{article}
\usepackage{amssymb,amsmath, amsbsy,epsfig, amsthm,textcomp}
\usepackage{a4wide,bigints,times,cite,bbm}

\newcommand{\squishlist}{
 \begin{list}{$\bullet$}
  { \setlength{\itemsep}{0pt}
     \setlength{\parsep}{3pt}
     \setlength{\topsep}{3pt}
     \setlength{\partopsep}{0pt}
     \setlength{\leftmargin}{1.5em}
     \setlength{\labelwidth}{1em}
     \setlength{\labelsep}{0.5em} } }

\newcommand{\squishend}{
  \end{list}  }

\usepackage{hyperref}
\hypersetup{
     colorlinks   = true,
     citecolor    = blue,
     linkcolor    = blue,
}
\newcommand{\bs}{\boldsymbol}

\newcommand{\IGNORE}[1]{}

\begin{document}
\title{A note on integrating products of linear forms over the unit simplex}
\author{Giuliano Casale\\Department of Computing\\Imperial College London, UK\\{\sf g.casale@imperial.ac.uk}}
\date{\today}
\maketitle

\begin{abstract}
Integrating a product of linear forms over the unit simplex can be done in polynomial time if the number of variables $n$ is fixed~\cite{Bal11}. In this note, we highlight that this problem is equivalent to obtaining the normalizing constant of state probabilities for a popular class of Markov processes used in queueing network theory. In light of this equivalence, we survey existing computational algorithms developed in queueing theory that can be used for exact integration. For example, under some regularity conditions, queueing theory algorithms can exactly integrate a product of linear forms of total degree $N$ by solving $N$ systems of linear equations.
\end{abstract}

\section{Introduction}
Let $\Delta=\{\bs x \in \mathbb{R}^n | x_i\geq 0, \sum_{i=1}^n x_i=1\}$ be the unit simplex and denote by $dm$ the integral Lebesgue measure on $\Delta$. Denote by $\theta_1,\theta_2,\ldots,\theta_d$ a collection of linear forms on $\mathbb{R}^n$, where $\theta_{ij}$ is the $i$th element of $\theta_j$ and define the coefficient matrix $\bs \theta=(\theta_{ij}) \in \mathbb{R}^{nd}$. Also let $N_1$,$\ldots$,$N_d$ be a set of nonnegative integers, with $N=\sum_{i=1}^d N_i$ and $\bs N=(N_1,\ldots,N_d)$.

Recently, \cite{Bal11} considers the problem of integrating a homogeneous polynomial function $f$ of degree $N$ over $\Delta$, observing that this can be reduced to computing a finite collection of integrals of the type
\begin{equation}
\label{eq:simplex}
J({\bs \theta},\bs N)=\bigintsss_\Delta\, \prod_{j=1}^d \left(\sum_{i=1}^n \theta_{ij} x_i\right)^{N_j} dm
\end{equation}
A polynomial-time algorithm to compute \eqref{eq:simplex} is then proposed in \cite{Bal11}, which obtains this integral by determining the coefficient of $z_1^{N_1}\cdots z_d^{N_d}$ in the Taylor expansion of
\begin{equation}
\label{eq:genfun}
T(z_1,\ldots,z_d)=\frac{1}{\prod_{i=1}^{n} (1-\sum_{j=1}^d z_j\theta_{ij})}
\end{equation}
in $O(N^d)$ time. Our goal is to examine alternative methods to compute \eqref{eq:simplex} that arise from queueing network theory~\cite{BolGMT06}.

We first observe that \eqref{eq:genfun} corresponds to the generating function of the {\em normalizing constant} $G({\bs \theta},\bs N)$ of state probabilities for a class of finite Markov processes known as {\em product-form closed queueing networks}, which have been extensively used in performance evaluation of computer and communication systems~\cite{BasCMP75,HarC02}. We provide in Appendix A a brief introduction to this class of Markov models; additional background may be found in \cite{Lav89,BolGMT06}.

The connection between normalizing constants and \eqref{eq:simplex} is explicitly noted in \cite{Cas17}, which shows that
\begin{equation}
\label{eq:GIeq}
J({\bs \theta},\bs N) = \frac{N_1!N_2!\cdots N_d!}{(N+n-1)!} G({\bs \theta},\bs N)
\end{equation}
Therefore, algorithms developed in queueing theory to compute $G({\bs \theta},\bs N)$ may be readily used to integrate products of linear forms over the unit simplex. 

The rest of this note reviews exact computational methods for $G({\bs \theta},\bs N)$. Although in queueing theory it is assumed that $\theta_{ij}\geq 0$, computational algorithms developed therein for $G({\bs \theta},\bs N)$ do not rely on this assumption and can therefore be applied also to linear forms with arbitrary values of $\theta_{ij}$. 

In the following sections, we denote by $\bs N-\bs 1_{j}$ the vector obtained from $\bs N$ by decreasing its $j$th element $N_j$ by one unit. We also indicate with $\bs \theta - \theta_j$ the matrix obtained from $\bs \theta$ by removing a row with elements $\theta_{j1},\ldots,\theta_{jd}$, and similarly denote by $\bs \theta +\theta_j$ the matrix obtained by appending to $\bs \theta$ a new row with elements $\theta_{j1},\ldots,\theta_{jd}$.

\section{Exact computational algorithms}

\subsection{Recurrence relations}
Recurrence relations are the standard method used in queueing theory to compute $G({\bs \theta},\bs N)$. Existing methods differ for computational requirements and ease of implementation. We here review two classic methods, convolution and RECAL, and the method of moments, a class of algorithms that scales efficiently under large degree $N$. We point to \cite{HarC02,HarL04,ChaS80,Lam82,Lam83} for other recursive algorithms.

\subsubsection{Convolution algorithm}
The convolution algorithm is a recurrence relation for normalizing constants presented in its most general form in~\cite{ReiK75b}, extending the case $d=1$ first developed in \cite{Buz73}. The method relies on the recurrence relation
\begin{equation}
\label{eq:CA}
G({\bs \theta},\bs N) = G({\bs \theta}-\theta_{n},\bs N) + \sum_{j=1}^d \theta_{nj} G({\bs \theta},\bs N-\bs 1_{j})
\end{equation}
with termination conditions $G(\emptyset,\cdot)=0$, $G(\cdot,\bs 0)=1$, where $\bs 0=(0,\ldots,0)$, and $G(\cdot,\bs N-\bs 1_{j})=0$ if $N_j=0$. It can be verified that the algorithm requires ${\cal O}(N^d)$ time and space for fixed $n$ and $d$. Variants of this algorithm exist~\cite{ChaS80,Lam83}, for example a specialized version has been proposed for problems with sparse $\bs \theta$~\cite{Lam82}.

\subsubsection{Recurrence by chain algorithm (RECAL)}
Compared to convolution, the RECAL algorithm is more efficient on models with large $d$~\cite{ConG86}. The method is based on the recurrence relation
\begin{equation}
\label{eq:RECAL}
G({\bs \theta},\bs N) = N^{-1}_d \sum_{i=1}^n \theta_{id} G({\bs \theta}+\theta_{i},\bs N-\bs 1_d)
\end{equation}
with similar termination conditions as the convolution algorithm.  The computational complexity is ${\cal O}(N^n)$ time and space for fixed $n$ and $d$. The method is well-suited for parallel implementation and can be optimized for sparse $\theta_{ij}$ coefficients~\cite{GreM89}.

\subsubsection{Method of moments}
The method of moments simultaneously applies \eqref{eq:CA} and \eqref{eq:RECAL} to a set of normalizing constants that differ for the elements and composition of $\bs \theta$ and $\bs N$. Different rules exist to define this set of normalizing constants, called the {\em basis}, leading to multiple variants of the method~\cite{Cas09,Cas11a,Cas11b}. Assume that the normalizing constants in the basis are grouped in a vector $\bs g(\bs N)$, then the method of moments defines a matrix recurrence relation
\begin{equation}
\label{eq:mom}
\bs A(\bs \theta, \bs N)\bs g(\bs N)=\bs B(\bs \theta, \bs N)\bs g(\bs N-\bs 1_j)
\end{equation}
for any $1\leq j\leq d$ and where $\bs g(\bs 0)$ can be computed explicitly from the termination conditions of \eqref{eq:CA}-\eqref{eq:RECAL}. Here $\bs A(\bs \theta, \bs N)$ and $\bs B(\bs \theta, \bs N)$ are square matrices, with constant or decreasing sizes as the recurrence progresses, depending on the implementation. The coefficients within these matrices depend on $j$. Thus, to avoid to recompute these matrices at each step, the method of moments first performs a recursion along dimension $j=d$, then along $j={d-1}$, and so forth up to reaching $\bs N=\bs 0$. An explicit example of the basic method may be found in \cite{Cas08}. Provided that $\bs A(\bs \theta, \bs N)$ is invertible at all steps of the recursion, it is possible to solve \eqref{eq:mom} and obtain $G(\bs \theta,\bs N)$ from the elements of $\bs g(\bs N)$. A necessary condition for invertibility is that $\theta_{ir}\neq \theta_{jr}$, $\forall i,j,r$.

Compared to existing algorithms, \eqref{eq:mom} finds $G(\bs \theta,\bs N)$ after solving $N$ systems of linear equations. Since the order of $\bs A(\bs \theta, \bs N)$ does not depend on $N$, the theoretical complexity of the method is ${\cal O}(N)$ time and ${\cal O}(1)$ space for fixed $n$ and $d$. However, implementation complexity is larger due to the cost of exact algebra, which is normally required to solve \eqref{eq:mom} without error propagation~\cite{Cas06}.


\subsection{Explicit solutions}
\subsubsection{Case $d=1$}
In this section we consider the case $d=1$, where we can simplify notation by indicating $\bs N$ with $N$ and $\theta_{i1}$ with $\theta_i$.
It has been known since long time that in the case $d=1$, and provided that all $\theta_{i}$ coefficients are distinct, the normalizing constant can be written as~\cite{Koe58}
\begin{equation}
\label{eq:Koe58}
G({\bs \theta},N) = \sum_{i=1}^{n} \frac{\theta^{N+n-1}_{i}}{\prod_{k\ne i}(\theta_{i}-\theta_{k})}
\end{equation}
that is simply the divided difference $[\theta_{1},\ldots,\theta_{n}] x^{N+n-1}$. It is useful to note that in the case where some coefficients $\theta_{ij}$ coincide, \eqref{eq:Koe58} generalizes to~\cite{BerM93}
\begin{align}
G({\bs \theta},N) &= [\theta_{i},\ldots,\theta_{n}] x^{N+n-1} \nonumber\\
                      &= \sum_{j=1}^{n'} (-1)^{m_j-1} \theta_{j}^{N+n-m_j}\sum_{\substack{\bs r\geq 0\\r=m_j-1}} (-1)^{r_j} {N+r_j \choose r_j} \prod_{k=1\atop k\neq j}^{n'} {m_k+r_k-1\choose r_k} \frac{\theta_{k}^{r_k}}{(\theta_{j}-\theta_{k})^{m_k+r_k}} \label{eq:gen}
\end{align}
where $n'\leq n$ is the number of distinct coefficients $\theta_{i}$, and $m_j$ denotes the number of coefficients identical to $\theta_{j}$.

\subsubsection{Case $d>1$}
For arbitrary values of $d$, \cite{Cas17} derives some generalizations of \eqref{eq:Koe58}. The first one is
\begin{equation}
\label{eq:explicit1}
G({\bs \theta},\bs N) = \sum_{\bs 0 \leq \bs t\leq \bs N} \frac{(-1)^{N-t}}{N_1!\cdots N_d!} \prod_{j=1}^d {N_j \choose t_j} \sum_{i=1}^{n} \frac{(\sum_{j=1}^d t_j\theta_{ij})^{N+n-1}}{\prod_{k\ne i}(\sum_{j=1}^d t_j(\theta_{ij}-\theta_{kj}))}
\end{equation}
where $\bs t=(t_1,\ldots,t_d)$, $t=\sum_{j=1}^d t_j$, which has the same ${\cal O}(N^{d})$ time complexity of the convolution algorithm, but ${\cal O}(1)$ space requirements. This expression holds assuming that $\theta_{ij}$ values for given $j$ are all distinct, otherwise the following generalized expression based on \eqref{eq:gen} should be used~\cite{Cas17}
\begin{multline}
G({\bs \theta},\bs N) = \sum_{\bs 0\leq \bs t\leq \bs N} \frac{(-1)^{N-t}}{N_1!\cdots N_d!}\prod_{r=1}^d{N_r \choose t_r} \sum_{j=1}^{n'} (-1)^{m_j-1} (\textstyle\sum_{s=1}^dt_s\theta_{js})^{N+n-m_j}\\
\times\sum_{\substack{\bs r\geq 0\\r=m_j-1}} (-1)^{r_j} {N+r_j \choose r_j} \prod_{k=1\atop k\neq j}^{n'} {m_k+r_k-1\choose r_k} \frac{(\sum_{s=1}^dt_s\theta_{ks})^{r_k}}{(\sum_{s=1}^dt_s(\theta_{js}-\theta_{ks}))^{m_k+r_k}}
\end{multline}
Another explicit expression is given by
\begin{equation}
\label{eq:explicit2}
G({\bs \theta},\bs N) = \sum_{\bs h\geq \bs 0:\atop h\leq N} \frac{(-1)^{N-h}}{N_1!\cdots N_d!} {N+n-1 \choose N-h} \prod_{j=1}^d \left(\sum_{i=1}^n h_i \theta_{ij}\right)^{N_j}
\end{equation}
where $\bs h=(h_1,\ldots,h_n)$, $h=\sum_{i=1}^n h_i$. This expression has ${\cal O}(N^{n+1})$ time and ${\cal O}(1)$ space requirements.

It is also worth noting that since the integrand of $\bs J(\bs \theta,\bs N)$ is a polynomial, the cubature rules proposed in \cite{GruM78} provide alternative explicit expressions for $G({\bs \theta},\bs N)$, which become exact for a sufficiently high interpolation degree, corresponding to a ${\cal O}((N/2)^{d})$ time complexity.
We point to \cite{Gor90,Ger95,BerM93} for other works on explicit expressions for the normalizing constant.


\section{Conclusion}
In this note, we have highlighted a connection between queueing network theory and the integration of products of linear forms over the unit simplex as in \eqref{eq:simplex}. In order to solve queueing network models, a normalizing constant is required, which can be computed with the recurrence relations and the explicit expressions that we have reviewed. This normalizing constant readily provides the value of the integral \eqref{eq:simplex}.

While the scope of the present note is restricted to exact methods, numerical approximations and asymptotic expansions are also available for $G({\bs \theta},\bs N)$, such as \cite{ChoLW95,WanCS16,KneT92,KogY96,Kog01}.

\section{Acknowledgement}
The author thanks Yosef Rinott for helpful comments on an earlier version of this manuscript.

\bibliographystyle{abbrv}

\begin{thebibliography}{10}

\bibitem{Bal11}
V.~Baldoni, N.~Berline, J.~A.~d. Loera, M.~K{\"o}ppe, and M.~Vergne.
\newblock How to integrate a polynomial over a simplex.
\newblock {\em Mathematics of Computation}, 80:297--325, 2011.

\bibitem{BasCMP75}
F.~Baskett, K.~M. Chandy, R.~R. Muntz, and F.~G. Palacios.
\newblock Open, closed, and mixed networks of queues with different classes of
  customers.
\newblock {\em JACM}, 22:248--260, 1975.

\bibitem{BerM93}
A.~Bertozzi and J.~McKenna.
\newblock Multidimensional residues, generating functions, and their
  application to queueing networks.
\newblock {\em SIAM Review}, 35(2):239--268, 1993.

\bibitem{BolGMT06}
G.~Bolch, S.~Greiner, H.~de~Meer, and K.~S. Trivedi.
\newblock {\em Queueing Networks and Markov Chains}.
\newblock Wiley, 2006.

\bibitem{Buz73}
J.~P. Buzen.
\newblock Computational algorithms for closed queueing networks with
  exponential servers.
\newblock {\em Comm. of the ACM}, 16(9):527--531, 1973.

\bibitem{Cas06}
G.~Casale.
\newblock An efficient algorithm for the exact analysis of multiclass queueing networks with large population sizes.
\newblock {\em Proc. of ACM SIGMETRICS}, 169--180, 2006.

\bibitem{Cas08}
G.~Casale.
\newblock An application of exact linear algebra to capacity planning models.
\newblock {\em ACM Communications in Computer Algebra}, 42(4):202--205, 2008.

\bibitem{Cas09}
G.~Casale.
\newblock {CoMoM}: Efficient class-oriented evaluation of multiclass
  performance models.
\newblock {\em IEEE Trans. on Software Engineering}, 35(2):162--177, 2009.

\bibitem{Cas11b}
G.~Casale.
\newblock A generalized method of moments for closed queueing networks.
\newblock {\em Performance Evaluation}, 68(2):180--200, 2011.

\bibitem{Cas11a}
G.~Casale.
\newblock Exact analysis of performance models by the method of moments.
\newblock {\em Performance Evaluation}, 68(6):487--506, 2011.

\bibitem{Cas17}
G.~Casale.
\newblock Accelerating performance inference over closed systems by asymptotic
  methods.
\newblock {\em Proc. ACM Meas.
Anal. Comput. Syst (POMACS)}, 1(1), 2017. To be presented at ACM SIGMETRICS 2017, preprint available at {\small\url{https://spiral.imperial.ac.uk/handle/10044/1/43431}}.

\bibitem{ChaS80}
K.~M. Chandy and C.~H. Sauer.
\newblock Computational algorithms for product-form queueing networks models of
  computing systems.
\newblock {\em Comm. of the ACM}, 23(10):573--583, 1980.

\bibitem{ChoLW95}
G~L. Choudhury, K.~K. Leung, and W.~Whitt.
\newblock Calculating normalization constants of closed queuing networks by
  numerically inverting their generating functions.
\newblock {\em JACM}, 42(5):935--970, 1995.

\bibitem{ConG86}
A.~E. Conway and N.~d. Georganas.
\newblock {\sc RECAL} - {A} new efficient algorithm for the exact analysis of
  multiple-chain closed queueing networks.
\newblock {\em Journal of the ACM}, 33(4):768--791, 1986.

\bibitem{Ger95}
A.~I. Gerasimov.
\newblock On normalizing constants in multiclass queueing networks.
\newblock {\em Oper. Res.}, 43(4):704--711, 1995.

 \bibitem{Gor90}
 J.~J. Gordon.
 \newblock The evaluation of normalizing constants in closed queueing networks.
 \newblock {\em Oper. Res.}, 38(5):863--869, 1990.

\bibitem{GreM89}
A.~G. Greenberg and J.~McKenna.
\newblock Solution of closed, product form, queueing networks via the {RECAL}
  and {TREE}-{RECAL} methods on a shared memory multiprocessor.
\newblock {\em ACM SIGMETRICS Performance Evaluation Review
  (PER)}, 17(1):127--135, 1989.

\bibitem{GruM78}
A.~Grundmann, H.M.~M\"oller.
\newblock Invariant integration formulas for the n-simplex by combinatorial
  methods.
\newblock {\em SIAM J. on Numerical Analysis}, 15(2):282--290, 1978.

\bibitem{Har85}
P.~G. Harrison.
\newblock On normalizing constants in queueing networks.
\newblock {\em Operations Research}, 33(2):464--468, 1985.

\bibitem{HarL04}
P.~G. Harrison and Ting Ting Lee
\newblock A New Recursive Algorithm for Computing Generating Functions in Closed Multi-Class Queueing Networks.
\newblock {\em Proc. of IEEE MASCOTS}, 231--238, 2004.

\bibitem{HarC02}
P.~G. Harrison and S.~Coury.
\newblock On the asymptotic behaviour of closed multiclass queueing networks.
\newblock {\em Performance Evaluation}, 47(2):131--138, 2002.

\bibitem{KneT92}
C.~Knessl, C.~Tier.
\newblock Asymptotic expansions for large closed queueing networks with
  multiple job classes.
\newblock {\em IEEE Trans. Computers}, 41(4):480--488, 1992.

\bibitem{Koe58}
E.~Koenigsberg.
\newblock Cyclic queues.
\newblock {\em Operational Research Quarterly}, 9, 1:22--35, 1958.

\bibitem{Kog01}
Y.~Kogan.
\newblock Asymptotic expansions for probability distributions in large loss and
  closed queueing networks.
\newblock {\em Perform. Eval. Rev.}, 29(3):25--27, Dec. 2001.

 \bibitem{KogY96}
 Y.~Kogan, A.~Yakovlev.
 \newblock Asymptotic analysis for closed multichain queueing networks with
   bottlenecks.
 \newblock {\em QUESTA}, 23:235--258, 1996.

\bibitem{Lam82}
S.~Lam.
\newblock Dynamic scaling and growth behavior of queueing network normalization
  constants.
\newblock {\em Journal of the ACM}, 29(2):492--513, 1982.

\bibitem{Lam83}
S.~Lam.
\newblock A simple derivation of the {\sc mva} and {\sc lbanc} algorithms from
  the convolution algorithm.
\newblock {\em IEEE Trans. on Computers}, 32:1062--1064, 1983.

\bibitem{Lav89}
S.~S. Lavenberg.
\newblock A perspective on queueing models of computer performance.
\newblock {\em Performance Evaluation}, 10(1):53--76, 1989.

\bibitem{ReiK75b}
M.~Reiser and H.~Kobayashi.
\newblock Queueing networks with multiple closed chains: Theory and
  computational algorithms.
\newblock {\em IBM J. Res. Dev.}, 19(3):283--294, 1975.

\bibitem{WanCS16}
W.~Wang, G.~Casale, C.~Sutton.
\newblock A Bayesian Approach to Parameter Inference in Queueing Networks.
\newblock {\em ACM TOMACS}, 27(1), 2016.

\end{thebibliography}

\appendix

\section{Product-form closed queueing networks}

A foundational result in queueing theory is the BCMP theorem~\cite{BasCMP75}, which establishes that networks of $n$ queueing stations with $d$ job classes admit a product-form equilibrium distribution when each station operates under one of four scheduling disciplines: First-Come First-Served (FCFS) with class-independent exponential service times, Processor Sharing (PS), Last-Come First-Served with Preemptive Resume (LCFS-PR), or Infinite Server (IS). Under these conditions, the equilibrium state probability factorizes across stations as
\begin{equation}
\label{eq:bcmp}
\mathbb{P}(\bs k) = \frac{1}{G(\bs \theta, \bs N)} \prod_{i=1}^n f_i(\bs k_i)
\end{equation}
where $\bs k_i=(k_{i1},\ldots,k_{id})$ is the vector of jobs at station $i$, $f_i(\bs k_i)$ is a function determined by the scheduling discipline and service rates at station $i$, and $G(\bs \theta, \bs N)$ is a normalizing constant ensuring that probabilities sum to one. The BCMP theorem applies to open, closed, and mixed networks; in what follows, we focus on the closed case.

In closed queueing networks, a finite set of $N$ jobs circulate among a network of $n$ nodes, called queueing stations, according to a set of transition probabilities. Jobs are partitioned into $d$ types, called classes, and the network topology is closed, meaning that the number of jobs inside the network remains constant over time and equal to $N_j$ in class $j$. Upon visiting a station $i$, a job of class $j$ is processed at rate $\mu_{ij}=\theta^{-1}_{ij}k^{-1}_i$ when a total of $k_i$ jobs are simultaneously present at the station. Therefore, the larger the number of jobs at a station, the smaller the rate.

A common question that arises in these models is to determine the joint stationary distribution of the number $k_{ij}$ of class-$j$ jobs residing at station $i$ when the Markov process reaches equilibrium. The analytic form of this distribution is known to be~\cite{BasCMP75}
\begin{equation}
\label{eq:Pm}
\mathbb{P}(\bs k)= \frac{1}{G_{\bs \theta}(\bs N)}\prod_{i=1}^n \frac{k_i!}{\prod_{j=1}^d k_{ij}!} \prod_{l=1}^d \theta^{k_{il}}_{il}
\end{equation}
where $\bs k\in {\cal S}$ is a state vector, ${\cal S}$ is a state space given by
\[
{\cal S}=\Biggl\{\bs k\in \mathbb{R}^{nd} \;\Big|\Bigr.\; k_{ij}\geq 0,\; \sum_{i=1}^n k_{ij}=N_j\Biggr\}
\]
and the normalizing constant is obtained by requiring that $\sum_{\bs k\in {\cal S}} \mathbb{P}(\bs k)=1$, {\em i.e.},
\begin{equation}
\label{eq:G}
G({\bs \theta},\bs N)=\sum_{\bs k\in {\cal S}} \prod_{i=1}^n \frac{k_i!}{\prod_{j=1}^d k_{ij}!} \prod_{l=1}^d \theta^{k_{il}}_{il}
\end{equation}

\end{document}